\documentclass{article}
\usepackage{amsmath}

\begin{document}

\begin{titlepage}

\title{ Relativistic Landau and oscillator levels in a symmetric gauge field
in a non-commutative complex space }
\author{S. Zaim and H. Rezki \\
D\'{e}partment de Physique, Facult\'{e} des Sciences de la Mati\`{e}re,\\
Universit\'{e} Hadj Lakhdar - Batna 1, Algeria}
\date{}
\maketitle
\thispagestyle{empty}
\begin{abstract}
In this work we obtain the exact solution for relativistic Landau problem
plus oscillator potential in a complex symmetric gauge field in a non-commutative
complex space, using the algebraic techniques of creation and
annihilation operators. It is shown that the relativistic Landau problem in
a complex symmetric gauge field in a non-commutative complex space is
similar behavior to the Pauli equation in the presence of a symmetric gauge
field in a commutative ordinary space. We derive the exact non-commutative
Landau and oscillator energy levels, while the non-relativistic limit of the
energy spectrum is obtained. We show that the energy is not degenerate and
is splitted into two levels, as in the Zeeman effect.
\end{abstract} 

\begin{quote}
\textbf{\sc Pacs numbers}: 02.40.Gh, 03.65.Ge, 03.65.Fd
\end{quote} \vspace*{4cm}
\end{titlepage}
\section{\protect\bigskip Introduction}

\quad There is a lot of interest in recent years on the study of non-commutative
canonical-type, quantum mechanics, quantum field theory and string theory $%
\left[ 1,2,3\right] $. On the other hand, the solutions of classical dynamical
problems of physical systems obtained in terms of complex space variables
are well-known. There are also interests on the complex quantum mechanical
systems (in two dimensions) $\left[ 4,5,6\right] $, in which we consider a
quantum relativistic Landau problem and harmonic oscillator in
non-commutative complex space $\left[ 7\right] $ (so coordinate and momentum
operators of this space are written as $\hat{z}=\hat{x}+i\hat{y}$ and $p_{\hat{z%
}}=\left( p_{x}-ip_{y}\right) /2$), where
\begin{equation}
\hat{x}^{\mu }=x^{\mu }-\frac{\theta ^{\mu \nu }}{2}p_{\nu }.
\end{equation}

For the non-commutative space of canonical-type, the parameter $\theta ^{\mu
\nu }$ is an anti-symmetric real matrix of length-square dimension. It
appears that the most natural places to search the non-commutative complex
coordinates effects are non-relativistic and relativistic quantum mechanics systems
in two dimensions. So far many interesting topics in
non-commutative non-relativistic and relativistic quantum mechanics in two
dimensions such as oscillator in the presence of uniform magnetic field $%
\left[ 8-25\right] $, and Landau problem $\left[ 26-31\right] $, have been
studied extensively. The purpose of this paper is to study the relativistic
Landau problem and relativistic oscillator in the presence of uniform complex
magnetic field on non-commutative complex space, where the non-commutative
complex coordinates could give an additional contribution. In this work, we
apply the algebraic techniques of creation and annihilation operators
to solve the relativistic Landau problem and relativistic oscillator in the
presence of uniform magnetic field on non-commutative complex space. In this
formalism, the presence of energy levels degeneracy was completely removed.

This paper is organized as follows: In section 2, the Landau problem in
non-commutative complex space is exactly solved, the corresponding exact
energy levels and algebraic function states are obtained respectively, while
the non-relativistic limit of the energy spectrum is obtained. In section 3, the
Klein-Gordon eigenvalues equation of oscillator in the presence of symmetric
complex gauge field is exactly solved in non-commutative complex space. The
conclusion is given in Section 4.

\section{ Landau problem in Non-commutative complex space}

\quad The conditions under which the relativistic Landau problem in non-commu-\\tative
quantum complex space and the Pauli equation are equivalent theories are
explored. In two-dimensional space, the complex coordinates system $\left( z,%
\bar{z}\right) $ and momentum $\left( p_{z},\bar{p}_{z}\right) $ are defined
by:\emph{\ }%
\begin{eqnarray}
z &=&x+iy,\quad \bar{z}=x-iy, \\
p_{z} &=&\frac{1}{2}\left( p_{x}-ip_{y}\right) ,\quad \bar{p}_{z}=\frac{1}{2}%
\left( p_{x}+ip_{y}\right) =-p_{\bar{z}}
\end{eqnarray}
We are interested in introducing the non-commutative complex operators of
coordinates and momentum in a two-dimensional space:
\begin{eqnarray}
\hat{z}=\hat{x}+i\hat{y}=z+i\theta \bar{p}_{z}, &\qquad &\widehat{\bar{z}}=%
\hat{x}-i\hat{y}=\bar{z}-i\theta p_{z}, \\
\hat{p}_{z}=p_{z}=-i\frac{d}{dz}, &\qquad &\widehat{\bar{p}}_{z}=\bar{p}%
_{z}=i\frac{d}{d\bar{z}}
\end{eqnarray}%
The non-commutative algebra $(1)$ can be rewritten as:
\begin{equation}
\left[ \hat{z},\hat{\bar{z}}\right] =2\theta ,\,\left[ \hat{z},p_{\hat{z}}%
\right] =\left[ \hat{\bar{z}},p_{\hat{z}}\right] =0,\,\left[ \hat{z},p_{\hat{%
z}}\right] =\left[ \hat{\bar{z}},p_{\hat{z}}\right] =\hbar ,\,\left[ p_{\hat{%
z}},p_{\hat{z}}\right] =0.
\end{equation}
Now we will discuss the relativistic Landau problem on non-commutative
complex quantum space in this formulation, we consider an electron of charge
$e$ and mass $m$ moves on a complex space in the presence of a symmetric gauge
complex potential $A\left( i\frac{B}{2}z,-i\frac{B}{2}\bar{z}\right) $, the
relativistic quantum equation in complex space is defined by the following
form:
\begin{equation}
\left( 2\bar{p}_{z}-eA_{z}\right) \left( 2p_{z}-e\bar{A}_{z}\right) \psi
=\left( E^{2}-m^{2}\right) \psi .
\end{equation}%
which can be written in commutative complex space as:
\begin{equation}
\left( 4\bar{p}_{z}p_{z}+\frac{e^{2}B^{2}}{4}z\bar{z}-eB\left(
L_{z}+1\right) \right) \psi =\left( E^{2}-m^{2}\right) \psi .
\end{equation}
where $L_{z}=i\left( zp_{z}-\bar{z}\bar{p}_{z}\right) ,$ is the $z$-component
of the orbital angular momentum, then the Hamiltonian of the
system is given by:

\begin{equation}
H=\frac{2}{m}p_{z}p_{\bar{z}}+m\frac{\omega _{c}^{2}}{2}z\bar{z}-\omega
_{c}\left( L_{z}+1\right) ,\text{ \ \ \ \ }\omega _{c}=\frac{eB}{2m}
\end{equation}
In a non-commutative complex space, eq.(7) is described by the following
equation:
\begin{equation}
\left(
\begin{array}{cc}
\left( 2p_{z}-e\widehat{\bar{A}}_{z}\right) \left( 2\bar{p}_{z}+e\widehat{A}%
_{z}\right)  & 0 \\
0 & \left( 2\bar{p}_{z}+e\widehat{A}_{z}\right) \left( 2p_{z}-e\widehat{\bar{%
A}}_{z}\right)
\end{array}%
\right) \psi =\left( E^{2}-m^{2}\right) \psi .
\end{equation}
Using the definition of the non-commutative complex coordinates, we can
rewrite this equation in a commutative complex space as:

\begin{equation}
\left( \frac{2}{\tilde{m}}p_{z}p_{\bar{z}}+\frac{\tilde{m}}{2}\tilde{\omega}%
^{2}z\bar{z}-\frac{eB}{2m}L_{z}-s_{z}\frac{e^{2}B^{2}}{4m}\theta -\frac{%
e^{2}B^{2}}{8m}\theta L_{z}\right) \psi =\bar{E}\psi ,
\end{equation}
\bigskip
where $\tilde{m}=m\left( 1+\frac{eB}{2}\theta \right) $, $\tilde{%
\omega}=\frac{eB}{2\tilde{m}}\left( 1+\frac{eB}{4}\theta \right) ,s_{z}=\pm
1/2$ and $\bar{E}=\frac{E^{2}-m^{2}+eB}{2m}$.
We note that the term $s_{z}\frac{e^{2}B^{2}}{4m}\theta ,$ is similar to the
spin-magnetic momentum interaction and the term $\frac{e^{2}B^{2}}{8\tilde{m}%
}\theta L_{z},$ is similar to the spin-orbit interaction. So the equation (11)
is similar to the equation of the electron with spin $\frac{1}{2}$ in a plane
under a symmetric gauge field. Thus, the corresponding Hamiltonian of
equation (11) is written as:

\begin{equation}
H=\frac{2}{\tilde{m}}p_{z}p_{\bar{z}}+\tilde{m}\frac{\tilde{\omega}^{2}}{2}z%
\bar{z}-\omega _{c}\left( 1+\frac{eB}{4}\theta \right) L_{z}-s_{z}\frac{%
e^{2}B^{2}}{4m}\theta
\end{equation}
So that a critical point is obtained when the coefficient of $L_{z}$ equals to
zero in this case where the non-commutative parameter $\theta =-\frac{4}{eB}$. In
this critical point the Hamiltonian of the system is:

\begin{equation}
H=\frac{1}{2\tilde{m}}p^{2}+\frac{m}{2}\left( \frac{eB}{2m}\right)
^{2}r^{2}+2\frac{e}{2m}Bs_{z},
\end{equation}
where it represents the oscillation of single electron with spin $\frac{1}{2}$
in a constant magnetic field, where the energy spectrum is given by:

\begin{equation}
E^{2}=2eB\left( n\pm \frac{1}{2}\right) +m^{2}
\end{equation}
The non-relativistic limit is given as:

\begin{equation}
E_{nr}=\frac{eB}{m}\left( n\pm \frac{1}{2}\right) \text{ \ \ \ with \ \ }%
n=0,1,2,...
\end{equation}
Each of these energy levels is splitting into two levels, hence we can say
that the particle in non-commutative complex space describes the electron
with spin $1/2$ in magnetic field. Where the non-commutativity creates
automatically the total magnetic momentum of particle with spin $1/2$, which in
turnshifted creates the spectrum of energy.
If $\theta \neq -\frac{4}{eB},$ the equation $(11)$ can be written according
to the eigenvalues equation as following:

\begin{equation}
H\psi =\bar{E}\psi
\end{equation}
To solve this equation, we can use the algebraic techniques of creation and
annihilation operators. To this aim, we define:

\begin{eqnarray}
a &=&\frac{2ip_{z}+e\tilde{B}\bar{z}}{2\sqrt{e\tilde{B}}}, \\
b &=&\frac{-2i\bar{p}_{z}+e\tilde{B}z}{2\sqrt{e\tilde{B}}},
\end{eqnarray}%
where $\tilde{B}=\frac{B}{2}\left( 1+\frac{eB}{4}\theta \right) $, the
corresponding creation operators $a^{+}$ and $b^{+}$ satisfy the usual
commutation relations:

\begin{equation}
\left[ a,a^{+}\right] =\left[ b,b^{+}\right] =1.
\end{equation}%
All the other commutation relations are zero. Now we can write the equation $(16)$ in
terms of the operators $a^{+}a$ and $b^{+}b$ as:

\begin{equation}
\left( \frac{e\tilde{B}}{\tilde{m}}\left( b^{+}b+a^{+}a+1\right) +\frac{e%
\tilde{B}}{m}+\frac{e\tilde{B}}{m}\left( b^{+}b-a^{+}a\right) +s_{z}\frac{%
e^{2}B^{2}}{4m}\theta \right) \psi ^{\sigma _{z}}=\bar{E}\psi ^{\sigma _{z}},
\end{equation}
where $\sigma _{z}=\pm 1$, the states $\psi ^{\sigma _{z}}$ are labeled
by the number $n_{1}$ for the quanta excitation of the operator $a$, and the number $%
n_{2}$ for the quanta excitation of the operator $b$:

\begin{eqnarray}
a^{+}a\psi _{n_{1},n_{2}}^{\sigma _{z}} &=&n_{1}\psi _{n_{1},n_{2}}^{\sigma
_{z}} \\
b^{+}b\psi _{n_{1},n_{2}}^{\sigma _{z}} &=&n_{2}\psi _{n_{1},n_{2}}^{\sigma
_{z}}
\end{eqnarray}
The energy levels of the equation $(16)$ are given as follows:

\begin{eqnarray}
E^{2} &=&m^{2}+2\frac{me\tilde{B}}{\tilde{m}}\left( n_{2}+n_{1}+1\right) +2e%
\tilde{B}\left( n_{2}-n_{1}\right)-eB\pm \frac{e^{2}B^{2}}{4}\theta
\end{eqnarray}
The non-relativistic limit is given as:

\begin{equation}
E_{nr}=\frac{e\tilde{B}}{\tilde{m}}\left( n_{2}+n_{1}+1\right) +\frac{e%
\tilde{B}}{m}\left( n_{2}-n_{1}\right) -\frac{eB}{2m}\pm \frac{e^{2}B^{2}}{8m%
}\theta
\end{equation}
In this formulation, there is an important observation about the removed degeneracy of this
spectrum and its splitting into two levels. Such effects are similar to
the Zeeman splitting in a commutative space. We find that four
eigenstates components for the $n^{th}$ Landau levels with the quantum number
$\sigma _{z}$ have the form:

\begin{equation}
\psi _{n_{1},n_{2}}^{\sigma _{z}}=\left\vert n_{1},n_{2}\right\rangle
\left\vert \pm \right\rangle
\end{equation}
where

\begin{eqnarray}
\psi _{0,0}^{\sigma _{z}} &=&\left\vert 0,0\right\rangle \left\vert \pm
\right\rangle =\left\vert 0\right\rangle \left\vert \pm \right\rangle \\
\psi _{n_{1},0}^{\sigma _{z}} &=&\frac{\left( a^{+}\right) ^{n_{1}}}{\sqrt{%
n_{1}!}}\left\vert 0\right\rangle \left\vert \pm \right\rangle \\
\psi _{0,n_{2}}^{\sigma _{z}} &=&\frac{\left( b^{+}\right) ^{n_{1}}}{\sqrt{%
n_{2}!}}\left\vert 0\right\rangle \left\vert \pm \right\rangle \\
\psi _{n_{1},n_{2}}^{\sigma _{z}} &=&\frac{\left( a^{+}\right)
^{n_{1}}\left( b^{+}\right) ^{n_{1}}}{\sqrt{n_{1}!n_{2}!}}\left\vert
0\right\rangle \left\vert \pm \right\rangle
\end{eqnarray}
and

\begin{eqnarray}
s_{z}\psi _{n_{1},n_{2}}^{\sigma _{z}} &=&\frac{1}{2}\sigma _{z}\psi
_{n_{1},n_{2}}^{\sigma _{z}} \\
a^{+}a\psi _{n_{1},n_{2}}^{\sigma _{z}} &=&n_{1}\psi _{n_{1},n_{2}}^{\sigma
_{z}} \\
b^{+}b\psi _{n_{1},n_{2}}^{\sigma _{z}} &=&n_{2}\psi _{n_{1},n_{2}}^{\sigma
_{z}}
\end{eqnarray}
There are four independent states:\\
$$\psi _{n_{1},0}^{+}\left( z,\bar{z}%
\right),\psi _{n_{1},0}^{-}\left( z,\bar{z}\right) $$
and\\
 $$\psi_{0,n_{2}}^{+}\left( z,\bar{z}\right),\psi _{0,n_{2}}^{-}\left( z,\bar{z}%
\right)$$
These particles are positioned in the four equivalent points $%
\left( z^{+},\bar{z}^{+},z_{-},\bar{z}_{-}\right)$. These results came from the
fact that the particle has a spin $\frac{1}{2}$ induced by the non-commutativity
effects in complex space.

\section{ Relativistic Landau problem plus oscillator potential in non-commutative
complex space }
\quad We consider the movement of an electron by oscillation on the complex space $\left( z,%
\bar{z}\right) ,$ subjected to a complex gauge potential field $A\left( -ie%
\frac{B}{2}\bar{z},i\frac{B}{2}z\right) ,$ where $B$ is the magnetic
momentum. In this gauge, the relativistic quantum equation in complex space
can be defined by the following equation:
\begin{equation}
\left[ \left( 2p_{z}+ie\frac{B}{2}\bar{z}\right) \left( 2p_{\bar{z}}-ie\frac{%
B}{2}z\right) +m^{2}\omega ^{2}z\bar{z}\right] \psi =\left(
E^{2}-m^{2}\right) \psi ,
\end{equation}%
which can be rewritten as:

\begin{equation}
\left( 4p_{\bar{z}}p_{z}+\left( m^{2}\omega ^{2}+\frac{e^{2}B^{2}}{4}\right)
z\bar{z}-eB\left( L_{z}+1\right) \right) \psi =\left( E^{2}-m^{2}\right)
\psi .
\end{equation}
\bigskip The corresponding Hamiltonian of the equation (34) is:

\begin{equation}
H=\frac{2}{m}p_{z}p_{\bar{z}}+\frac{m}{2}\left( \omega ^{2}+\omega
_{c}^{2}\right) z\bar{z}-\omega _{c}\left( L_{z}+1\right)
\end{equation}
The eigenvalues for the Hamiltonian in equation $(35)$ are:
\begin{equation}
E^{2}=2m\left( \omega ^{2}+\omega _{c}^{2}\right) ^{1/2}\left(
n_{1}+n_{2}+1\right) +eB\left( n_{1}-n_{2}-1\right) +m^{2}
\end{equation}
The non-relativistic limit is given as:

\begin{equation}
E_{nr}=\left( \omega ^{2}+\omega _{c}^{2}\right) ^{1/2}\left(
n_{1}+n_{2}+1\right) +\omega _{c}\left( n_{1}-n_{2}-1\right)
\end{equation}
In the non-commutative complex space $\left( \hat{z},\hat{\bar{z}}\right) $,
the relativistic quantum oscillator in a complex symmetric gauge potential
field $A=\left( i\frac{B}{2}z,-i\frac{B}{2}\bar{z}\right)$ is described by
the following equation:
\begin{eqnarray}
\left(
\begin{array}{cc}
\left( 2p_{z}+ie\frac{B}{2}\hat{\bar{z}}\right) \left( 2p_{\bar{z}}-ie\frac{B%
}{2}\hat{z}\right) +m^{2}\omega ^{2}\hat{\bar{z}}\hat{z} & 0 \\
0 & \left( 2p_{\bar{z}}-ie\frac{B}{2}\hat{z}\right) \left( 2p_{z}+ie\frac{B}{%
2}\hat{\bar{z}}\right) +m^{2}\omega ^{2}\hat{z}\hat{\bar{z}}%
\end{array}%
\right) \psi  \notag \\=\left( E^{2}-m^{2}\right) \psi
\end{eqnarray}%
Using the relations $(3$) and $(4)$ we can rewrite the equation $(38)$ in
commutative complex space as:

\begin{eqnarray}
\left( 4(1-e\frac{B}{4}\theta )^{2}p_{z}p_{\bar{z}}+\left( m^{2}\omega
^{2}+\left( e\frac{B}{2}\right) ^{2}\right) z\bar{z}-2\left( e\frac{B}{2}%
\right) L_{z}+1\right.   \notag \\
\left. -\left( m^{2}\omega ^{2}+\left( e\frac{B}{2}\right) ^{2}\right)
\theta \left( L_{z}\mp 1\right) \right) \psi =\left( E^{2}-m^{2}\right) \psi
\end{eqnarray}%
The equation $(39)$ can be written in a very simple way as:
\begin{equation}
\left( \frac{2}{\tilde{m}}p_{z}p_{\bar{z}}+\frac{\tilde{m}}{2}\varpi ^{2}z%
\bar{z}-\left( \frac{\tilde{m}}{2}\varpi ^{2}\theta +\omega _{c}\right)
L_{z}+s_{z}m\varpi ^{2}\theta \right) \psi =\bar{E}\psi ,
\end{equation}
where $\varpi ^{2}=\frac{\omega ^{2}+\left( \frac{eB}{2m}\right) ^{2}}{(1+e%
\frac{B}{2}\theta )}$, $\tilde{m}=m(1+e\frac{B}{2}\theta )$ and $\bar{E}=%
\frac{E^{2}-m^{2}+eB}{2m}$. The equation $\left( 40\right) $ is similar to
the Pauli equation of motion for a fermion of spin $\frac{1}{2}$ in a
constant magnetic field. So that a critical point is obtained when the
coefficient of a new constant equals to zero, in this case the relation
between the magnetic momentum and the non-commutative parameter is given by:

\begin{equation}
B=-\frac{m^{2}\omega ^{2}}{2e}\theta
\end{equation}
The negative sign means that the non-commutative parameter is in the opposite
direction of the vector $\overrightarrow{L}_{z}.$ If we substitute the
parameter $\theta $ in Eq. $(40)$, it leads to an oscillator with spin $%
\frac{1}{2}$ on a commutative complex space in a constant magnetic field:

\begin{equation}
\left( \frac{2}{m}p_{z}p_{\bar{z}}+\frac{m}{2}\left( \omega ^{2}+\left(
\frac{eB}{2m}\right) ^{2}\right) z\bar{z}-2\frac{\omega ^{2}+\left( \frac{eB%
}{2m}\right) ^{2}}{m^{2}\omega ^{2}}s_{z}B\right) \psi =\bar{E}\psi ,
\end{equation}
The accompanying Hamiltonian to the equation is written as:

\begin{equation}
H=\frac{2}{m}p_{z}p_{\bar{z}}+\frac{m}{2}\left( \omega ^{2}+\left( \frac{eB}{%
2m}\right) ^{2}\right) z\bar{z}-2\frac{\omega ^{2}+\left( \frac{eB}{2m}%
\right) ^{2}}{m^{2}\omega ^{2}}s_{z}B
\end{equation}
The eigenvalues for the Hamiltonian in equation $(43)$ are:

\begin{equation}
E^{2}=2m\left( \omega ^{2}+\left( \frac{eB}{2m}\right) ^{2}\right)
^{1/2}\left( 2n+1\right) \mp 2\frac{\omega ^{2}+\left( \frac{eB}{2m}\right)
^{2}}{m\omega ^{2}}B-eB+m^{2}
\end{equation}
The non-relativistic limit is given as:

\begin{equation}
E_{nr}=\left( \omega ^{2}+\left( \frac{eB}{2m}\right) ^{2}\right)
^{1/2}\left( 2n+1\right) -\frac{eB}{2m}\mp \frac{\omega ^{2}+\left( \frac{eB%
}{2m}\right) ^{2}}{m^{2}\omega ^{2}}B
\end{equation}
At the critical point, the energy spectrum is splitting to two levels. So
the charged oscillator in non-commutative complex space at the critical point
is similar to Pauli particle in commutative ordinary space.
If $B\neq -\frac{m^{2}\omega ^{2}}{2e}\theta ,$ the equation $(40)$ can be
written according to the eigenvalues equation as following:

\begin{equation}
H\psi =\bar{E}\psi
\end{equation}
where

\begin{equation}
H=\frac{2}{\tilde{m}}p_{z}p_{\bar{z}}+\frac{\tilde{m}}{2}\varpi ^{2}z\bar{z}%
-\left( \frac{\tilde{m}}{2}\varpi ^{2}\theta +\omega _{c}\right)
L_{z}+s_{z}m\varpi ^{2}\theta
\end{equation}
To solve Eq.$(46)$, we can use the algebraic techniques of creation and
annihilation operators. For this purpose, we define:

\begin{eqnarray}
\tilde{a} &=&\frac{2ip_{z}+\tilde{m}\varpi \bar{z}}{2\sqrt{\tilde{m}\varpi }}%
, \\
\tilde{b} &=&\frac{-2i\bar{p}_{z}+\tilde{m}\varpi z}{2\sqrt{\tilde{m}\varpi }%
},
\end{eqnarray}
\bigskip We can now write the equation $(46)$ in terms of these operators as:
\begin{equation}
\left( \varpi \left( \tilde{a}^{+}\tilde{a}+\tilde{b}^{+}\tilde{b}+1\right)
-\left( \frac{m}{2}\varpi ^{2}\theta +\omega _{c}\right) \left( \tilde{a}^{+}%
\tilde{a}-\tilde{b}^{+}\tilde{b}\right) +s_{z}m\varpi ^{2}\theta \right. %
\bigg)\psi _{n_{1},n_{2}}^{\sigma _{z}}=\bar{E}\psi _{n_{1},n_{2}}^{\sigma
_{z}},
\end{equation}
where the states $\psi _{n_{1},n_{2}}^{\sigma _{z}}$ are labeled by the
number $n_{1}$ for the quanta excitation of the operator $\tilde{a}$,
and the number $n_{2}$ for the quanta excitation of the operator $\tilde{b}$:

\begin{eqnarray}
\tilde{a}^{+}\tilde{a}\psi _{n_{1},n_{2}}^{\sigma _{z}} &=&n_{1}\psi
_{n_{1},n_{2}}^{\sigma _{z}} \\
\tilde{b}^{+}\tilde{b}\psi _{n_{1},n_{2}}^{\sigma _{z}} &=&n_{2}\psi
_{n_{1},n_{2}}^{\sigma _{z}}
\end{eqnarray}
The energy eigenvalues for eq. $(50)$ are given by:
\begin{equation}
E^{2}=m^{2}-eB+2m\varpi \left( n_{1}+n_{2}+1\right) -2m\left( \frac{m}{2}%
\varpi ^{2}\theta +\omega _{c}\right) \left( n_{1}-n_{2}\right) \pm
m^{2}\varpi ^{2}\theta ,
\end{equation}
The non-relativistic limit is given as:

\begin{equation}
E_{nr}=\frac{\varpi }{m}\left( n_{1}+n_{2}+1\right) -\left( \frac{m\varpi
^{2}\theta +2\omega _{c}}{2m}\right) \left( n_{1}-n_{2}\right) -\frac{eB}{2m}%
\pm \frac{m}{2}\varpi ^{2}\theta
\end{equation}
where $n_{1},n_{2}=0,1,2,....$ and $m_{l}=n_{1}-n_{2}=0,\pm 1,\pm 2,.......$\\
The energy level is splitting into two levels ( they are labeled by $%
n_{1},n_{2} $), which removed the degeneracy. It is similar to the Zeeman effect.
Hence we can say that the particle in non-commutative complex space describes
the particle with spin $1/2$ in magnetic field, therefore the system with
spin in a magnetic field will have a resonance $\left[ 32\right] $. Then the
critical values of $\theta =-\frac{2e}{m^{2}\omega ^{2}}$ can be considered
as a resonance point. At this point, the system can be treated as landau problem
with spin $ 1/2$.

\section{Conclusion}

\quad In this work we started from charged relativistic quantum particle and
charged oscillator in a uniform magnetic field in a canonical
non-commutative complex space. By using the Moyal product up to first order
in the non-commutative parameter $\theta $, we derived the deformed
relativistic Landau problem and Klein-Gordon oscillator equations. By
solving them exactly we found that the energy removed the degeneracy and
shifted up it to the first order in $\theta $ by two levels, such effects are
similar to the Zeeman splitting in a commutative ordinary space. In
addition, we also obtained the non-relativistic limit of the energy spectrum.


\begin{thebibliography}{99}
\bibitem{1} V.P. Nair and A.P. Polychronakos, Phys.Lett. B505, 267(2001).

\bibitem{2} P. K. Ghosh, Eur. Phys. J. C42,355 (2005).

\bibitem{3} P. D. Alvarez, J. Gomis, K. Kamimura and M. S.Plyushchay, Phys.
Lett. B659, 906 (2008).

\bibitem{4} C. Duval, P. A. Horv\`{a}thy, Phys.Lett.B479:284-290,2000

\bibitem{5} J. Ben Geloun, S. Gangopadhyay, F. G Scholtz, Europhys. Lett.
86: 51001, 2009

\bibitem{6} A.I.Arbab ,EPL,98(2012)30008.

\bibitem{7} S.Zaim, Int. J. Theor. Phys, 53 , 6 , 2014-2023 (2014).

\bibitem{8} Nair, V.P.: Phys. Lett. B 505, 249 (2001)

\bibitem{9} Chaichian, M., et al.: Phys. Rev. Lett. 86, 2716 (2001)

\bibitem{10} Smailagic, A., Spallucci, E.: Phys. Rev. D, Part. Fields 65,
107701 (2002)

\bibitem{11} Smailagic, A., Spallucci, E.: J. Phys. A, Math. Theor. 35, L363
(2002)

\bibitem{12} Wei, G.F., Long, C.Y., Long, Z.W., et al.: Chin. Phys. C 32,
247 (2008)

\bibitem{13} Gamboa, J., Loewe, M., Rojas, J.C.: Phys. Rev. D 64, 067901
(2001)

\bibitem{14} Wei, G.F., Long, C.Y., Long, Z.W., et al.: Chin. Phys. C 32,
338 (2008)

\bibitem{15} Muthukumar, B., Mitra, P.: Phys. Rev. D, Part. Fields 66,
027701 (2002)

\bibitem{16} Gamboa, J., et al.: Int. J. Mod. Phys. A 17, 2555 (2002).

\bibitem{17} Guo, G.J., Long, C.Y., Yang, Z.H.: Can. J. Phys. 87, 989 (2009)

\bibitem{18} Li, K., Wang, J.H., Dulat, S.: Int. J. Theor. Phys. 49, 134
(2010)

\bibitem{19} Mirza, B., Mohadesi, M.: Commun. Theor. Phys. 42, 664 (2004)

\bibitem{20} \ Deng, M.Q., Xu, Z.J., Wei, G.F.: J. At. Mol. Phys. 27, 365
(2010)

\bibitem{21} Wang, J.H., Li, K., Dulat, S.: Chin. Phys. C 32, 803 (2008)

\bibitem{22} Falek, M., Merad, M.: Commun. Theor. Phys. 50, 587 (2008)

\bibitem{23} Mirza, B., Narimani, R., Zare, S.: Commun. Theor. Phys. 55, 405
(2011)

\bibitem{24} Yang, Z.H., Long, C.Y., Qin, S.J., et al.: Int. J. Theor. Phys.
49, 644 (2010)

\bibitem{25} Yongjun Xiao \textperiodcentered\ Zhengwen Long
\textperiodcentered\ Shaohong Cai, Int J Theor Phys (2011) 50:3105--3111.

\bibitem{26} Pulak Ranjan Giri, P. Roy, Eur. Phys. J. C (2008) 57: 835--839.

\bibitem{27} F Delduc et al 2008 J. Phys.: Conf. Ser. 103 012020.

\bibitem{28} B.K.Pal, B. Roy, B. Basu, Phys. Lett. A 374:4369-4374 (2010).

\bibitem{29} S. Gangopadhyay, A. Saha, A. Halder, Phys. Lett. A
379:2956-2969 (2015).

\bibitem{30} S. Dulat, K. Li, Chin Phys. C32: 92-95 (2008).

\bibitem{31} J. Gamboa, M. Loewe, F. Mendez; J. C. Rojas, Mod. Phys. Lett. A
16: 2075-2078 (2001).

\bibitem{32} M. Mohadesi, B. Mirza, Commun.Theor.Phys. 42 (2004) 664-668.
\end{thebibliography}
\end{document}